\documentclass[a4paper,11pt]{article}
\usepackage{pos}

\usepackage{cancel}
\usepackage{amsmath} 
\usepackage{color} 

\definecolor{green}{rgb}{0,.5,0}

\definecolor{red}{rgb}{1,0,0}

\usepackage{graphicx}
\usepackage[subfigure]{graphfig}
\usepackage{epsfig}
\usepackage{slashed}
\usepackage{dcolumn}
\usepackage{latexsym}
\usepackage{multirow}

\title{The calculations of Nucleon Electric Dipole Moment using background field on Lattice QCD}

\author*[a]{Fangcheng He}
\author[b]{Michael Abramczyk}
\author[b]{Tom Blum}
\author[c,d]{Taku Izubuchi}
\author[e]{Hiroshi Ohki}
\author[a]{Sergey Syritsyn}

\affiliation[a]{
  Department of Physics and Astronomy, Stony Brook University, \\ 
  Stony Brook, NY 11794, USA
}
\affiliation[b]{
  Physics Department, University of Connecticut, \\
  Storrs, Connecticut 06269, USA
}
\affiliation[c]{
  RIKEN/BNL Research Center, Brookhaven National Laboratory, \\
  Upton, NY 11973, USA
}
\affiliation[d]{
  Physics Department, Brookhaven National Laboratory, \\
  Upton, New York 11973, USA
}
\affiliation[e]{
  Department of Physics, Nara Women's University, Nara 630-8506, Japan \\
}

\emailAdd{fangcheng.he@stonybrook.edu}

\abstract{
Measurements of nucleon and nuclei Electric Dipole Moments (EDMs) play an important role in
probing CP violation and exploring physics beyond the Standard Model. 
We extract the neutron EDM by measuring the energy shift of the nucleon two-point correlation
function in the presence of a background field. 
The UV divergence of the topological charge density operator is mitigated using gradient flow, 
and the diffusion effect induced by the gradient flow process is included into the fit ansatz.
Our calculations were carried out on two 2+1 DWF fermion, Iwasaki, gauge field ensembles
generated by the RBC/UKQCD collaborations with inverse lattice spacing 1.73 GeV and pion masses
of about 340 and 420 MeV.
}

\FullConference{The 40th International Symposium on Lattice Field Theory (Lattice 2023)\\
July 31st - August 4th, 2023\\
Fermi National Accelerator Laboratory\\}

\newcommand{\CP}{{CP}}
\newcommand{\mcE}{{\mathcal{E}}}

\newcommand{\mcO}{{\mathcal{O}}}

\newcommand{\CPviol}{{\cancel{\rm CP}}}
\newcommand{\bea}{\begin{eqnarray}}
\newcommand{\eea}{\end{eqnarray}}
\newcommand{\be}{\begin{equation}}
\newcommand{\ee}{\end{equation}}
\newcommand{\ba}{\begin{array}}
\newcommand{\ea}{\end{array}}
\newcommand{\lp}{\left}
\newcommand{\rp}{\right}
\newcommand{\DN}[1]{\textcolor{red}{#1}}

\newcommand{\la}{\langle}
\newcommand{\ra}{\rangle}
\newcommand{\Slash}[1]{{\ooalign{\hfil/\hfil\crcr$#1$}}}
\newcommand{\tins}{\tau}
\newcommand{\tsep}{t_f}
\newcommand{\tgf}{t^\text{gf}}

\newcommand{\Tr}{{\mathrm{Tr}}}

\begin{document}
\maketitle

\section{Introduction}
The nucleon electric dipole moment (nEDM) is an important indicator of CP(T)-symmetry violation. 
The nEDM prediction from the Standard Model's CKM matrix for the neutron is
$\approx 10^{-31}$ [$e $ cm], which is much smaller than the upper-limit determined through
experiments on neutrons (ILL) \cite{Baker:2006ts} and $^{199}$Hg \cite{Graner:2016ses}.
The SM prediction is also too small to explain the matter-antimatter asymmetry
compared to what is required by the Sakharov conditions.
Hence, the measurement of the nEDM is important in the search for  physics beyond Standard Model.

Strong interactions may be a source of CP violation known as the QCD "$\theta$-term",
$S_{\theta} = \theta Q$, where $Q$ is the topological charge.
Its contributions to the nEDM can be studied systematically from first principles
only using lattice QCD.
Many efforts have been made to calculate the nEDM using Lattice
QCD~\cite{Shintani:2005xg, Berruto:2005hg, Aoki:2008gv, Guo:2015tla, Alexandrou:2015spa,
Shintani:2015vsx, Abramczyk:2017oxr,Dragos:2019oxn,Alexandrou:2020mds,Bhattacharya:2021lol,Liang:2023jfj}.
In earlier calculations, an incorrect definition of the electric dipole form factor $F_3$
resulted in significant mixing with the Pauli form factor~\cite{Abramczyk:2017oxr}.
After subtracting the mixing term, those nEDM results became much smaller, comparable with
phenomenology, and universally dominated by statistical noise even with unphysical heavy quark
masses.

Currently, most of nEDM calculations on a lattice use the traditional form factor method
in which nEDMs are extracted as electric dipole form factors (EDFF) $F_3(Q^2)$ from 
$\CP$-odd corrections to matrix elements of the quark vector current due to topological charge.
This form factor has to be extrapolated to the forward limit ($Q^2\to0$) to obtain the nEDM.
The other method is to calculate the EDM from the nucleon energy shift 
$\Delta E\propto d_N(2\vec S\cdot\vec E)$ in a uniform background electric field
\cite{Shintani:2006xr, Shintani:2008nt,Abramczyk:2017oxr}.
This method has significant advantages: no forward limit is required, there is no parity mixing
between $F_2$ and $F_3$, and one only needs to calculate the $\CP$-odd part of the two- instead
of three-point function.

We present our $\theta$-nEDM results using the background field method.
Details of this method were first described in Ref.~\cite{Izubuchi:2020ngl}.
The background field method is described in Sec.~\ref{sec:cp}, the results are presented in
Sec.~\ref{sec:num}, and discussed in Sec~\ref{sec:summary}.

\section{Electric dipole moment from background field method}
\label{sec:cp}

In this section, we discuss and compare methods to compute the nEDM on a lattice.
The form factor method has been described in multiple publications (see, e.g.,
Ref.\cite{Abramczyk:2017oxr} and references therein).
The EDFF is defined as
\begin{eqnarray}
\label{eqn:ff_cpviol}
\langle p^\prime,\sigma^\prime |J^\mu|p,\sigma
\rangle_{\CPviol}
&=& \bar{u}_{p^\prime,\sigma^\prime} 
   \big[F_1(Q^2) \gamma^\mu + \big(F_2(Q^2) 
  + iF_3(Q^2)\gamma_5\big) \frac{i\sigma^{\mu\nu}q_\nu}{2M_N} \big] u_{p,\sigma}\,
\end{eqnarray}
where $Q^2=-(p^\prime-p)^2$, 
and $F_{1,2,3}$ are the Dirac, Pauli, and electric dipole form factors.
The forward limit of the latter yields the nEDM, $F_3(0)=2m_nd_n$. 
Although the nucleon states in QCD vacuum with $\CP$-violation are no longer parity eigenstates,
it is crucial to ensure that their spinors satisfy the positive-parity Dirac equation with real-valued mass,
$(\Slash{p}-m_N) u_{p,\sigma} = 0$,
otherwise $F_3$ receives a spurious contribution from $F_2$~\cite{Abramczyk:2017oxr},
\begin{equation}\label{eq:FFrel}
\tilde F_3(Q^2) = F_3\cos(2\alpha) - F_2\sin(2\alpha),
\end{equation}
where $\alpha$ is the parity mixing of the lattice nucleon spinor due to $\CP$-violation that
can be extracted from the two-point function.

Another approach to calculate the nucleon EDM is the background field method introduced in
\cite{Shintani:2006xr, Shintani:2008nt, Izubuchi:2007rmy,Izubuchi:2020ngl}.
It has also been extended to analyze CP-even properties, such as the electric
polarizability and magnetic moments of the nucleon~\cite{Detmold:2009dx, Detmold:2010ts}.
Since the proton accelerates in an electric field, and requires a more sophisticated analysis,
we study only the neutron EDM in this work.
After including the background field, the Dirac equation for the parity-positive neutron
spinor $u_N$ in the rest frame ($p_N=(m_N,\vec0)$) becomes in Euclidean space~\cite{Abramczyk:2017oxr}
\begin{align}
\lp[ i \slashed{p} + m_N 
  -\big(\frac12 F_{\mu\nu}\sigma^{\mu\nu}\big)\frac{\kappa + i\zeta\gamma_5}{2m_N} \rp] \, u_{N}
  = \lp(\ba{cc} 
    m_N - \frac{(\kappa + i\zeta)\vec\mcE\cdot\vec\sigma}{2m_N} &   -E_N \\ 
    -E_N &   m_N + \frac{(\kappa-i\zeta)\vec\mcE\cdot\vec\sigma}{2m_N}
  \ea\rp) \, u_N  = 0\,, 
\end{align}  
where $E_N$ is the neutron energy, $\kappa,\zeta=F_{2,3}(0)$ are magnetic and electric dipole
moments, and $\vec\mcE$ is the Euclidean electric field.
To the order linear in $\kappa,\zeta$, the nucleon energy $E_N$ is
\begin{equation}
E_N^2 = m_N^2 - i \zeta (\vec\Sigma \cdot\vec\mcE) + O(\kappa^2, \zeta^2) \,,
\quad\text{ or }\quad
E_N = m_N - \frac{i\zeta}{2m_N}\,(\vec\Sigma \cdot\vec\mcE) + O(\kappa^2, \zeta^2)
\end{equation}
where $\zeta/(2m_N)=d_N$ is the electric dipole moment and
$\vec\Sigma=\text{diag}[\vec\sigma,\vec\sigma]$ is the spin operator.
Note that the linear part of the energy shift 
$\delta E=-\frac{i\zeta}{2m_N}\,(\vec\Sigma \cdot\vec\mcE)$ 
is imaginary because of the analytic continuation in electric field on a Euclidean lattice,
implying that the correlation function acquires a complex phase.
Expanding the path integral in $\theta\ll1$, one obtains the $\CP$-violating correction to the
nucleon correlation function in the background field $\mcE$ on one hand, and the $t$-linear
correction on the other:
\begin{equation}
C_{2pt,\mcE,\theta} 
\approx \la N(t) \bar{N}(0)\ra_\mcE -i\theta \la Q \, N(t) \bar{N}(0)\ra_\mcE
\propto e^{-E t} \big[1 - t\delta E + O(\theta^2)\big]\,.
\end{equation}
The nEDM $\zeta=F_3(0)$ can be extracted from nucleon correlators in an electric field
$\vec\mcE = \mcE_z\hat z$ as
\begin{equation}
\label{eqn:edm_estimator_sum}
\frac{\zeta}{\theta} = i\frac{2m_N}{\mcE_z} \frac{\delta E}{\theta}
= -\frac{2m_N}{\mcE_z} \frac{d}{dt} \,  
  \frac {\mathrm{Tr}\big[T^+_{S_z} \la Q \, N(t)\bar{N}(0)\ra_{\mcE_z}\big]}
        {\mathrm{Tr}\big[T^+ \la N(t)\bar{N}(0)\ra_{\mcE_z}\big]}\,,
\end{equation}
where $T^+=\frac12(1+\gamma_4)$ is the positive-parity projector, and $T^+_{S_z} =
T^+\cdot(1+\Sigma_z)$ is the  spin-$\hat z$ projector.

So far, we have ignored excited states and negative parity state.
In practice, a multi-state model can be fitted to the time dependence of the EDM
estimator~(\ref{eqn:edm_estimator_sum}).
On the other hand, the formula~(\ref{eqn:edm_estimator_sum}) resembles the ``summation'' method
of computing ground-state matrix elements.
This relation is made apparent by the Feynman-Hellman theorem (recently discussed in
Ref.~\cite{Bouchard:2016heu} in the context of lattice QCD), so the EDM can
alternatively be calculated from the matrix element of the local density of topological charge,
\begin{equation}
\label{eqn:edm_bgem_plateau}
\frac\zeta\theta = -\frac{2m_N}{\mcE_z} \la N_\uparrow |q| N_\uparrow\ra_{\mcE_z} 
\approx -\frac{2m_N}{\mcE_z} 
  \frac{\Tr \big[T^+_{S_z} \la N(\tsep) \, q(\tins) \, \bar N(0)\ra_{\mcE_z}\big]}
       {\Tr \big[T^+       \la N(\tsep) \, \bar N(0)\ra_{\mcE_z}\big]}\,.
\end{equation}
Thus, the problem is reduced from computing a 4-point function between the nucleon fields
$N,\,\bar N$, the global topological charge $Q=\int d^4x q(x)$, and the vector current 
$J_\mu=\bar \psi\gamma_\mu \psi$ to a correlator of $N,\,\bar N$ and the \emph{local topological
charge density} $q(x)$ in uniform field:
\begin{equation}
\label{eqn:edm_bgem_3pt}
C_3(t_f,\tau) = \Tr \big[T^+_{S_z} \la N(\tsep) \, q(\tins) \, \bar N(0)\ra_{\mcE_z}\big] 
  = \big[T^+_{S_z}\big]_{\beta\alpha} \sum_{\vec y,\vec z}
    \la N_\alpha(\tsep, \vec y)\, q(\tins, \vec z) \, \bar N_\beta(0) \ra\,.
\end{equation}
This point is crucial to reducing fluctuations in correlators with any operator that leads to
``disconnected'' Wick contractions: topological charge, Weinberg 3-gluon interaction, isoscalar
2-quark and 4-quark interactions, and so on.
In this paper we concentrate on the topological charge but the methodology can be readily
extended to these other $\CPviol$ interactions.

Chiral symmetry is important for calculations involving topological charge.
We use gauge configurations generated with $N_f=2+1$ flavors of domain wall fermions (DWF)~\cite{RBC:2014ntl}.
(Tab.~\ref{tab:ens}).
\begin{table}[ht!]
\centering
\caption{\label{tab:ens}
Gauge ensembles used in this work~\cite{RBC:2014ntl}.}
\begin{tabular}{c|cc|ccc}
\hline\hline
 & size & $m_l/m_s$ & $m_\pi\,MeV$ & $N_{cfg}$ \\
\hline
I24\_m010 & $24^3\times64$ & 0.01/0.04 & 432 & 1100\\
\hline
I24\_m005 & $24^3\times64$ & 0.005/0.04 & 340 & 1400\\
\hline\hline
\end{tabular}
\end{table}

Periodicity on a lattice in space and time requires quantization of abelian field flux
$Q_q E_k L_k L_4 = 2\pi n$, where $Q_u=\frac23$ and $Q_d=-\frac13$ are quark charges.
A uniform electric field on the lattice is introduced by multiplying the gauge links by the
$U(1)$ phase~\cite{Detmold:2009dx}, 
\begin{align}
A_{x,4}  = 
n_{k4} \Phi_{k4} x_k, 
\quad \quad 
A_{x,k}|_{x_k=L_k-1}  = 
-n_{k4} \Phi_{k4} L_kx_4, 
\end{align}  
where $n_{\mu\nu}$ is the number of quanta and 
$\Phi_{\mu\nu}=\frac{6\pi}{L_\mu L_\nu}$ is the unit of flux through plaquette ($\mu\nu$).
Such potentials result in a uniform electric field $\vec{E}=(n_{14} \Phi_{14}, n_{24} \Phi_{24}, n_{34} \Phi_{34})$.

\section{Numerical results
  \label{sec:num}}

To calculate nucleon correlators, we use all-mode-averaging (AMA)~\cite{Shintani:2014vja} by combining
high-precision and truncated-CG samples computed with low-mode deflation of the preconditioned
Dirac operator.
We also employ low-mode averaging whereby we approximate the full-volume average of nucleon
correlators from low modes and combine it with AMA samples to correct bias,
which have resulted in a $50\%$ reduction of statistical errors:
\begin{equation}
\la \mcO\ra \approx 
\la\mcO^\text{LMA}\ra_{V_4} 
+ \la\mcO^\text{approx} - \mcO^\text{LMA}\ra_{N_{approx}}
+ \la\mcO^\text{exact} - \mcO^\text{approx}\ra_{N_{exact}}\,.
\end{equation}
In addition, we average over spin orientations $2S_z=\pm1$ and electric fields
$\mcE_z=\pm1,\pm2$.

\begin{figure}[ht!] 
\centering
\includegraphics[width=.48\textwidth]{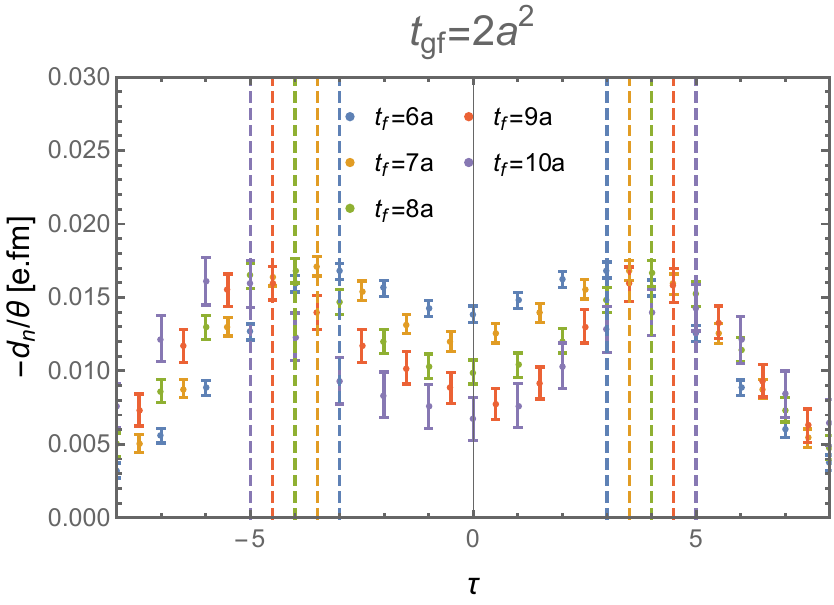}~
\includegraphics[width=.48\textwidth]{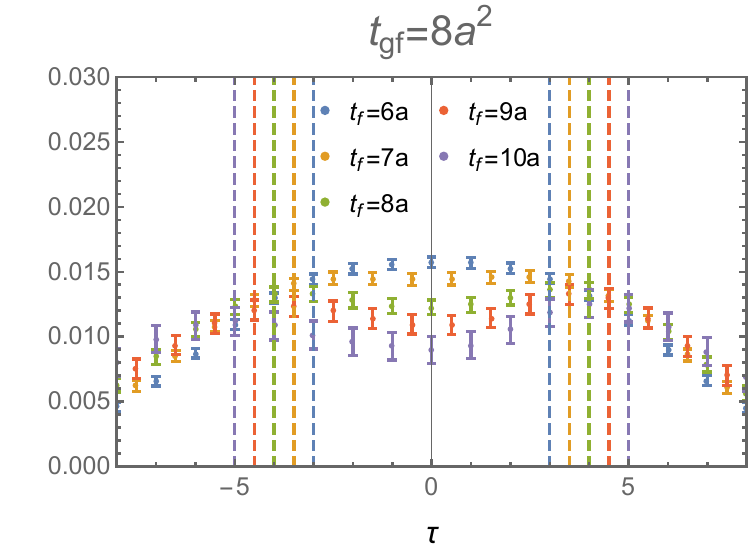}\\
\caption{Plateaus of neutron matrix elements of topological charge density converted to
EDM~(\ref{eqn:edm_bgem_plateau}), depending on the gradient flow time.
The vertical dashed lines of respective colors show the locations of the nucleon fields,
which lead to contact terms with the gluon operator on the same time slices.
As the gradient flow time increases, the contact terms ``diffuse'' into the EDM plateaus.}
\label{fig:gfdep}
\end{figure}

To use the Feynman-Hellman method, we need to determine the local topological charge density
$q(x)$.
We calculate $q(x)$ using 5-loop improved
discretization~\cite{deForcrand:1997esx} and gradient flow~\cite{Luscher:2010iy}.
Gradient flow helps suppress $O(a)$ size dislocations that contribute to fluctuation of the
global topological charge.
At large enough gradient flow time, the \emph{global topological charge} $Q$
becomes constant. 
However, the \emph{local density} (summed over the spatial volume) entering the correlator in
Eq.~(\ref{eqn:edm_bgem_plateau}), becomes delocalized (``diffused'') in the Euclidean time, 
which complicates analysis of its ground-state matrix elements.
This effect is visible in the plateaus for $d_n/\theta$~(\ref{eqn:edm_bgem_plateau}) shown
in Fig.~\ref{fig:gfdep} for source-sink separations $t=(6\ldots10)a$:
as the gradient flow time $\tgf$ is increased, the $\tau$-dependent features in the
ratios~(\ref{eqn:edm_bgem_plateau}) become more diffused.
In particular, it becomes hard to isolate the ground-state plateau from the contact term
contributions where the density $q(x)$ overlaps with the nucleon operators.

\begin{figure}[ht!]
  \centering
  \includegraphics[width=.45\textwidth]{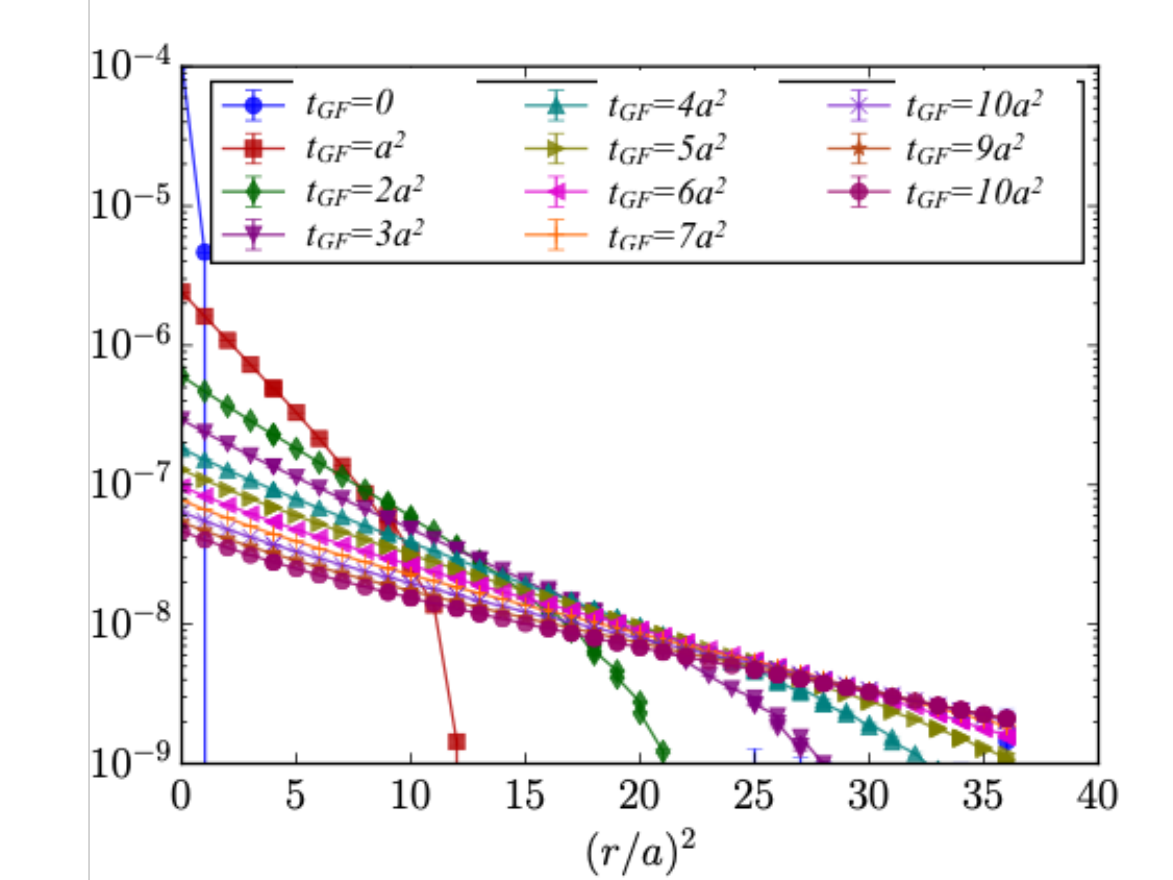}~
  \hspace{.04\textwidth}~
  \includegraphics[width=.45\textwidth]{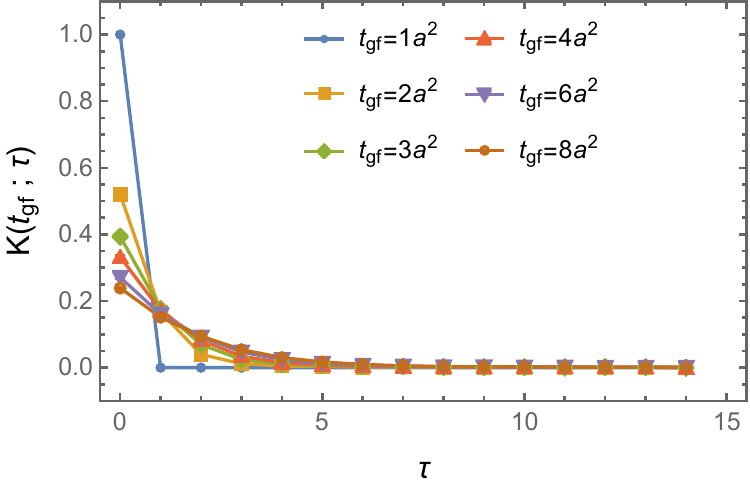}
  \caption{\label{fig:diffusion}
    (Left) correlation of the topological charge density $\la q(r) q(0)\ra$ 
    computed with $\tgf$ and
    (right) diffusion kernel $K(\tgf_2-\tgf_1;\tau)$~(\ref{eqn:qtopo_diff_kernel}).
  }
\end{figure}
We model this Euclidean-time diffusion effect of the gradient flow by 
\begin{equation}
q(\tgf_2;\tau)=\int dt' K(\tgf_2-\tgf_1;|\tau-\tau'|) \, q(\tgf_1;\tau')
\end{equation}
where $K(\tgf_2-\tgf_1;|\tau-\tau'|)$ is the diffusion kernel that can be extracted from
Fourier-transformed correlations of the topological charge 
\begin{equation}
\label{eqn:qtopo_diff_kernel}
K(\tgf_2-\tgf_1;\tau)
 = \frac1{L_t}\sum_k e^{i\omega_k \tau} 
      \sqrt{\frac{\tilde\chi(\tgf_2; \omega_k)}{\tilde\chi(\tgf_1; \omega_k)}}\,,
\quad
\tilde\chi(\tgf;\omega_k)= \sum_\tau e^{-i\omega_k\tau} 
  \langle q(\tgf;\tau) q(\tgf;0)\rangle\,,
\end{equation}
where the frequency $\omega_k = \frac{2\pi}{L_t}k$.
Numerical results for kernel $K(\tgf_2-\tgf_1;|\tau|)$ are shown in Fig.~\ref{fig:diffusion},
illustrating how the diffusion profile widens with increasing gradient flow time $\tgf$.
Similarly, the effect of ``diffusion'' on the three-point correlation
function~(\ref{eqn:edm_bgem_3pt}) can be written as
\begin{equation}
\label{eqn:c3pt_gf_convol}
\tilde{C_3}(\tgf;\tsep,\tins)
  = \sum_{\tau'} K(\tgf;|\tau-\tau'|)C_3(\tsep,{\tins}')\,,
\end{equation}
where $C_3(\tsep,\tau)$ is the correlation function that would be computed with a strictly local
definition of the topological charge $q(\tins)$.
Unfortunately, this diffusion mixes the neutron EDM signal with contributions from unwanted regions
$\tins\le0$ and $\tins\ge\tsep$, which we model as
\begin{equation}
 \label{eqn:c3pt_int_ext}
C_{3pt}(\tau,\tsep)=\lp\{\ba{ll}
  a_0 e^{-E_0\tsep}\lp(d_n +c_1 e^{-\Delta E_1\,\tau}+c_1e^{-\Delta E_1\,(\tsep-\tau)}
      +c_2e^{-\Delta E_1\,\tsep}\rp)\,, 
  & 0<\tau<\tsep\,,\\
  C_{ext}e^{-E_0 \tsep} e^{E_{ext}\tau}\,,
  & \tau\leq 0\,,\\ 
  C_{ext} e^{-E_0\tsep}e^{-E_{ext}(\tau-\tsep)} \,, 
  & \tsep\leq\tau\,.\\  
  \ea\rp.
\end{equation}
The first line corresponds to the matrix element where the gluon operator is located between the
source and sink.
The second and the third lines represent the contributions from contact terms ($\tau=0$ or
$\tau=\tsep$) and ``crossed'' regions with $q(\tins)$ located outside of the source-sink time interval.

\begin{figure}[ht!] 
\centering
\includegraphics[width=.48\textwidth]{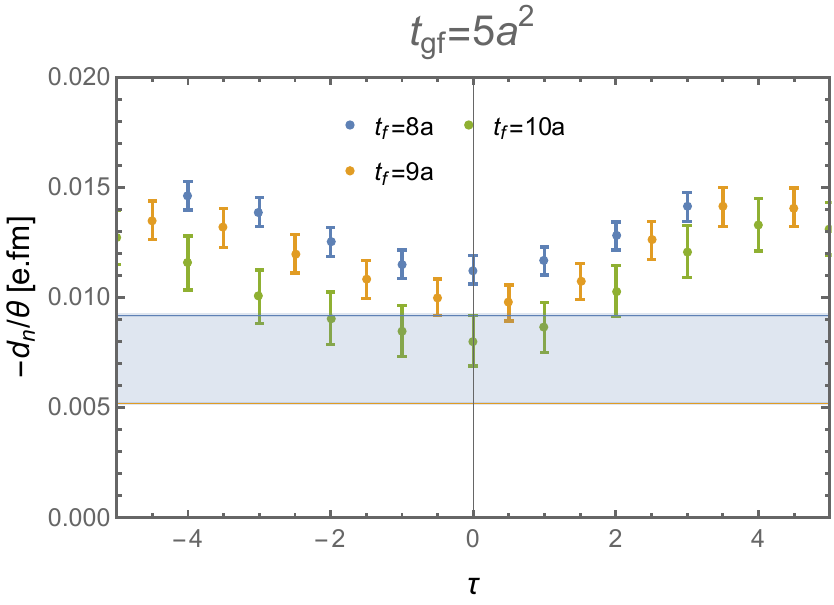}~
\includegraphics[width=.48\textwidth]{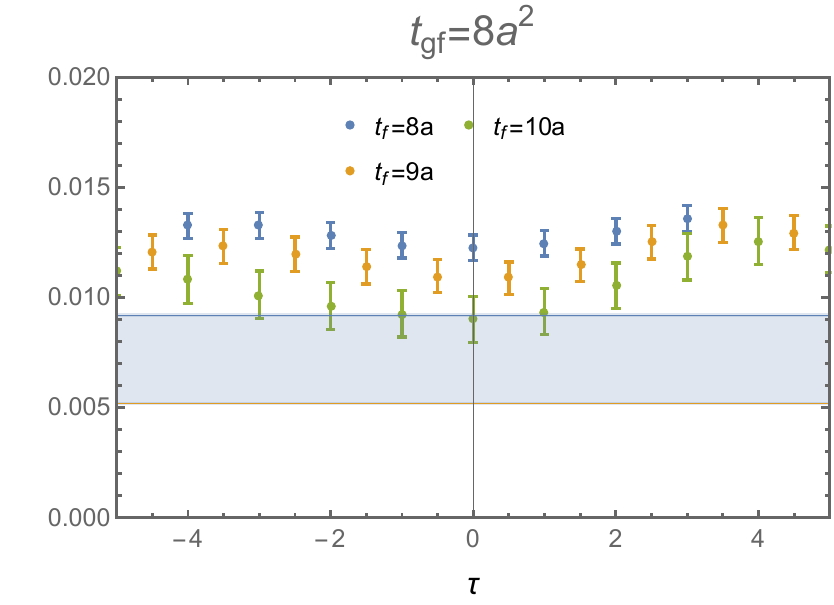}\\
\caption{\label{fig:3ptfit}
  Fits of combined Euclidean-time~(\ref{eqn:c3pt_int_ext}) and gradient flow-time 
  dependence~(\ref{eqn:c3pt_gf_convol}) of lattice nEDM correlation functions.
  The band is the fit result using Eq.~(\ref{eqn:c3pt_int_ext}).}
\end{figure}

We perform a fit of three-point~(\ref{eqn:edm_bgem_3pt}) data to ansatz~(\ref{eqn:c3pt_int_ext})
joint with a two-state fit of two-point neutron data to constrain ground- and excited-state
energies $E_0$, $E_0+\Delta E_1$.
For the former, we select data with with longer range $8a\le\tsep\le10a$ and $5a^2\le\tgf\le8a^2$.
The fit result for the ground-state EDM $d_n/\theta$ is shown in Fig.~(\ref{fig:3ptfit}).
We obtain the following values at the two pion masses
\begin{align*}
[d_n/\theta]_{340\,\text{MeV}} &= -0.0072(20)^\text{stat}\,,
\\
[d_n/\theta]_{432\,\text{MeV}} &= -0.0060(20)^\text{stat}\,.
\end{align*}

Using chiral prediction $d_n(m_\pi^2) \sim c_1 m_\pi^2+c_2 m_\pi^2 \log(m_\pi^2)$, we can obtain
only a tentative estimate of the value at the physical point
$(d_n/\theta)^\text{phys}_\text{lin.+log.} =-0.0036(21)^\text{stat}$.
The uncertainty of the extrapolation is substantially constrained by chiral symmetry in our
EDM calculation.

\section{Summary}
\label{sec:summary}

In this work, we have calculated the neutron EDM employing the background field method on
two gauge ensembles with chirally symmetric quarks and pion masses $\approx340$ and 420 MeV.
Unlike in the traditional form-factor approach, we calculate the forward matrix element of the
topological charge density in a simultaneously spin- and electrically-polarized nucleon state.
This method is instrumental to limit the growth of the stochastic noise from the global
topological charge as the physical volume of a lattice increases.
Without the need to extrapolate in the momentum transfer $Q^2\to0$, the systematic uncertainty
can be further dramatically reduced.

The main obstacle in using this method is the difficulty of determining topological charge
density.
In this work, we have used the field-theoretical definition from the gluon fields combined with
gradient flow.
However, gradient flow leads to ``diffusion'' in Euclidean time and mixing of the nEDM signal
with contact terms and contributions from $n\bar{n}$-pair states.
We have implemented the numerical procedure to extract the diffusion kernel from lattice data
and incorporate it in the fits of correlators of the nucleon and the topological charge density.
With only two pion mass points available, we could perform only a tentative estimate of the
physical-point value.
Another point at lighter pion mass $m_\pi\approx250$ MeV is currently being investigated.

We plan to explore the fermionic definition~\cite{Alexandrou:2020mds,Luscher:2021bog}, which may
be especially advantageous when combined with the low-mode approximation for neutron correlation
function.
Further, this method can be easily applied to other $\CPviol$ operators, and might be especially
beneficial for the nEDM induced by Weinberg's three-gluon operator and 4-quark $\CPviol$
operators.

\section*{Acknowledgments}
The research reported in this work made use of computing and long-term storage facilities provided by the
USQCD Collaboration, which are funded by the Office of Science of the U.S. Department of Energy,
and the Hokusai supercomputer of the RIKEN ACCC facility.
We are grateful for the gauge configurations provided by the RBC/UKQCD collaboration.
SS and FH are supported by the National Science Foundation under CAREER Award PHY-1847893.
Any opinions, findings, and conclusions or recommendations expressed in this material are those
of the author(s) and do not necessarily reflect the views of the National Science Foundation.
HO is supported in part by JSPS KAKENHI Grants No. 21K03554 and No. 22H00138. 
TB and MA were partially supported by the US DOE under the award DE-SC001033.
TI is supported by U.S. Department of Energy (DOE) under award DE-SC0012704, SciDAC-5 LAB 22-2580,
and also Laboratory Directed Research and Development (LDRD No. 23-051) of BNL and RIKEN BNL 
Research Center.

\bibliography{paper}
\bibliographystyle{aip_ep}

\end{document}